\documentclass[%
  fleqn,colorlinks,linkcolor=blue,citecolor=blue,urlcolor=blue]{eptcs}
\usepackage[latin1]{inputenc}
\usepackage{amsfonts}
\usepackage{amsmath}
\usepackage{amssymb}
\usepackage{amsthm}
\usepackage[T1]{fontenc}
\usepackage{newpxtext,newpxmath}
\usepackage{z-eves}
\usepackage{multicol}
\usepackage{algorithm}
\usepackage{algpseudocode}
\usepackage{comment}
\usepackage{listings}
\usepackage{xspace}
\usepackage{fancyvrb}
\usepackage{tabularx}
\usepackage{framed}
\usepackage{enumitem}
\usepackage{tikz}
\usetikzlibrary{positioning,fit,calc,babel,decorations.text,patterns,
                shapes,arrows,automata,decorations.pathmorphing,
                shadows.blur}

\newtheorem*{teorema*}{Teorema}

\newcommand{\card}[1]{\lvert #1 \rvert}

\newcommand{\setlog}{$\{log\}$\xspace}

\newcommand{\Size}{size}

\newcommand{\Rel}{rel}

\newcommand{\Dom}{dom}
\newcommand{\Comp}{comp}
\newcommand{\set}{set}
\renewcommand{\Cup}{un}
\renewcommand{\Cap}{inters}
\newcommand{\disj}{\parallel}
\newcommand{\Diff}{diff}
\newcommand{\In}{in\xspace}
\newcommand{\Disj}{disj}
\newcommand{\Subseteq}{subset}
\renewcommand{\Subset}{ssubset}
\newcommand{\Neq}{neq\xspace}

\newcommand{\Ncup}{nun}
\newcommand{\Ncap}{ninters}
\newcommand{\Ndiff}{ndiff}
\newcommand{\Nin}{nin\xspace}
\newcommand{\Ndisj}{ndisj}
\newcommand{\Nsubseteq}{nsubset}

\newcommand{\Ndom}{ndom}
\newcommand{\Ran}{ran}

\newcommand{\Inv}{inv}

\newcommand{\Dres}{dres}
\newcommand{\Rres}{rres}
\newcommand{\Ndres}{dares}
\newcommand{\Nrres}{rares}
\newcommand{\Rimg}{rimg}
\newcommand{\Oplus}{oplus}
\newcommand{\Apply}{apply}

\newcommand{\Pfun}{pfun}

\newcommand{\List}{slist}
\newcommand{\Head}{head}
\newcommand{\Tail}{tail}
\newcommand{\Last}{last}
\newcommand{\Front}{front}
\newcommand{\Add}{add}
\newcommand{\Concat}{concat}
\newcommand{\Filter}{filter}
\newcommand{\Extract}{extract}

\title{}
\date{}

\begin{document}

\thispagestyle{empty}
\begin{center}
\mbox{}
\vfill
{\bf
{\Huge \{\textit{log}\}} \\[3mm]
{\Huge Applications to Software \\[3mm]
 Specification, Prototyping and \\[3mm] Verification} \vfill
{\large Maximiliano Cristiá} \\
E-mail: \href{mailto:cristia@cifasis-conicet.gov.ar}{\nolinkurl{cristia@cifasis-conicet.gov.ar}} \\
Universidad Nacional de Rosario and CIFASIS \\
Argentina \\[5mm]
{\large Gianfranco Rossi} \\
E-mail: \href{mailto:gianfranco.rossi@unipr.it}{\nolinkurl{gianfranco.rossi@unipr.it}} \\
Universtià di Parma \\
Italy
}
\vfill
\mbox{}
\end{center}

\pagebreak

\thispagestyle{empty}
\begin{abstract}
\textsc{Abstract}\hspace{3mm}
This document shows how Z specifications can be translated into \setlog and, later, on how \setlog can be used to run simulations and automated proofs. This can help users of other specification languages such as B and VDM to use \setlog along the same lines. The presentation is rather informal and user-oriented. More technical and formal presentations can be found in the papers published by the authors. We also assume the reader has at least a basic knowledge of the Z notation.
\end{abstract}

\vfill

\tableofcontents

\pagebreak

\DefineShortVerb[commandchars=\\\$\$]{\@}

\section{Preliminaries -- Installing \setlog}
\setlog (`setlog') is a constraint logic programming language and a satisfiability solver based on set theroy. As such it can be used as an automated theorem prover. One of \setlog's distinctive features is that sets are first-class entities of the language.

\setlog was first developed by Gianfranco Rossi and his PhD students in Italy during the mid '90. Since 2012 Gianfranco Rossi and Maximiliano Cristiá work together in extending \setlog in several ways.

Technical presentations about \setlog and the theory behind it can be found in several academic articles \cite{DBLP:journals/jlp/DovierOPR96,Dovier00,DBLP:journals/jar/CristiaR20,DBLP:conf/RelMiCS/CristiaR18,DBLP:conf/cade/CristiaR17,jar-ris,CristiaRossiSEFM13,DBLP:journals/jar/CristiaR21,setlog-itp-cj,DBLP:journals/corr/abs-2009-00999}.

\bigskip

\setlog is a Prolog program. Then, you first need to install a Prolog interpreter to run \setlog. So far \setlog runs only on SWI-Prolog (\url{http://www.swi-prolog.org}). After installing SWI-Prolog you must download  \setlog and all the library files from \setlog's home page:

\vspace{3mm}
\centerline{\url{http://people.dmi.unipr.it/gianfranco.rossi/setlog.Home.html}}

\bigskip

This document is focused on showing how Z specifications can be translated into \setlog and, later, on how \setlog can be used to run simulations and automated proofs. This can help users of other specification languages such as B and VDM to use \setlog along the same lines. The presentation is rather informal and user-oriented. More technical and formal presentations can be found in the papers published by the authors. We also assume the reader has at least a basic knowledge of the Z notation.

\section{\label{traduccion}Translating Z specifications into \setlog}
Many Z specifications can be easily translated into \setlog. This means that \setlog can serve as a programming language in which a prototype of a Z specification can be immediately implemented. 

We will consider Z specifications of a very specific form. Here, we will show how these Z specifications can be translated into \setlog. To that end we will use a running example. Later on we will explain with some detail how Z elements not appearing in the example can be translated into \setlog; we will see that some Z elements can be translated in more than one way.

\subsection{The running example}
The specification to be used as running example is one of the classic Z specifications first used by Spivey in several articles and books \cite{Spivey00}. It's known as the \emph{birthday book}. It's a system which records people's birthdays, and is
able to issue a reminder when the day comes round.

\subsubsection{The Z specification}
In our account of the system, we need to deal with people's names and
with dates. Then, we introduce the following basic types.
\begin{zed}
[NAME,DATE]
\end{zed}

Now we can define the state schema of the specification as follows.
\begin{schema}{BirthdayBook}
known: \power NAME \\
birthday: NAME \pfun DATE
\end{schema}
where $known$ is the set of names with birthdays recorded; and $birthday$ is a function which, when applied to certain names, gives the birthdays associated with them.

The initial state of the birthday book is the following.
\begin{schema}{BirthdayBookInit}
BirthdayBook
\where
known = \emptyset \\
birthday = \emptyset
\end{schema}

The following schema describes the predicates that should be state invariants.
\begin{schema}{BirthdayBookInv}
BirthdayBook
\where
known = \dom birthday
\end{schema}
As can be seen, the value of $known$ can be derived from the value of $birthday$. This makes $known$ a
\emph{derived} component.  It would be possible to specify the system without mentioning $known$ at all.
However, giving names to important concepts helps to make specifications more
readable. The specification doesn't commit the programmer to represent $known$ explicitly in an implementation.

The first operation we specify is how to add a birthday to the birthday book. As always we first model the normal behavior, then the abnormal behaviors and finally we assemble all the schemas in a single schema expression.
\begin{schema}{AddBirthdayOk}
\Delta BirthdayBook \\
name?:NAME \\
date?:DATE
\where
name? \notin known \\
known' = known \cup \{name?\} \\
birthday' = birthday \cup \{name? \mapsto date?\}
\end{schema}

\begin{schema}{NameAlreadyExists}
\Xi BirthdayBook \\
name?:NAME
\where
name? \in known
\end{schema}

\begin{zed}
AddBirthday == AddBirthdayOk \lor NameAlreadyExists
\end{zed}

The second operation to be specified is the one that shows the birthday of a given person.
\begin{schema}{FindBirthdayOk}
\Xi BirthdayBook \\
name?:NAME \\
date!: DATE
\where
name? \in known \\
date! = birthday(name?)
\end{schema}

\begin{schema}{NotAFriend}
\Xi BirthdayBook \\
name?:NAME
\where
name? \notin known
\end{schema}

\begin{zed}
FindBirthday == FindBirthdayOk \lor NotAFriend
\end{zed}

Finally we have an operation listing all the persons whose birthday is a given date.
\begin{schema}{Remind}
\Xi BirthdayBook \\
today?:DATE \\
cards!: \power NAME
\where
cards! = \dom(birthday \rres \{today?\})
\end{schema}

\subsubsection{\label{codigosetlog}The \setlog code}
The \setlog code of the translation of the Z specification should be saved in a file with extension \Verb+.slog+. It is convenient to put this file in the same folder where \setlog was installed.

Most Z schemas are translated into \setlog \emph{clauses}. A \setlog clause is a sort of subroutine or subprogram or procedure of a regular programming language. Each clause can receive zero or more arguments. When a Z schema representing an operation is translated, the corresponding clause receives as arguments all the state variables, all the input variables and all the output variables.

In \setlog variables must always begin with an uppercase letter or the underscore character (\Verb+_+), although this is usually saved for special cases. Any identifier beginning with a lowercase letter is a constant. The prime character ($'$) cannot be used as part of a variable's name. Therefore, in order to denote the after state variables we will put the underscore character at the end (`\Verb+_+'). Then, for instance, the state variables of the birthday book will be \Verb+Known+ and \Verb+Birthday+; and those of the after state will be \Verb+Known_+ and \Verb+Birthday_+. In the same way, $?$ y $!$ cannot be part of variables' names in  \setlog. In these cases we will use \Verb+_i+ and \Verb+_o+ as suffixes denoting input and output variables, respectively. Then, for example, $name?$ (see schema $AddBirthdayOk$) becomes \Verb+Name_i+ in \setlog; and $cards!$ (see schema $Remind$) becomes \Verb+Cards_o+.

Hence, the interface of the \setlog clause corresponding to the Z operation named $AddBirthday$ is the following:
\begin{Verbatim}
addBirthday(Known,Birthday,Name_i,Date_i,Known_,Birthday_)
\end{Verbatim}
where \Verb+Name_i+ and \Verb+Date_i+ correspond to input variables $name?$ and $date?$ declared in $AddBirthday$. Note that althugh $AddBirthday$ begins with an uppercase letter \Verb+addBirthday+ begins with a lowercase letter because \setlog clauses must begin in that way; on the other hand, although $name?$ and $date?$ begin with lowercase letters, \Verb+Name_i+ and \Verb+Date_i+ begin with uppercase letters because we want them to variable arguments. Finally, \Verb+Known+ and \Verb+Birthday+ represent the before state while \Verb+Known_+ y \Verb+Birthday_+ represent the after state.

Now we can give the definition of the \Verb+addBirthday+ clause:
\begin{Verbatim}
addBirthday(Known,Birthday,Name_i,Date_i,Known_,Birthday_) :-
  addBirthdayOk(Known,Birthday,Name_i,Date_i,Known_,Birthday_)
  or
  nameAlreadyExists(BirthdayBook,Name_i,BirthdayBook_).
\end{Verbatim}
where \Verb+or+ denotes logical disjunction. Observe that \Verb+Date_i+ isn't passed to \Verb+nameAlreadyExists+ as an argument because it isn't declared in $NameAlreadyExists$; besides, look that the clause ends in a dot.

In turn the definition of \Verb+addBirthdayOk+ is the following:
\begin{Verbatim}
addBirthdayOk(Known,Birthday,Name_i,Date_i,Known_,Birthday_) :-
  Name_i nin Known &
  un(Known,{Name_i},Known_) &
  un(Birthday,{[Name_i,Date_i]},Birthday_).
\end{Verbatim}
where:
\begin{itemize}
\item The \Verb+&+ symbol denotes logical conjunction.
\item \Verb+nin+ corresponds to the $\notin$ operator. That is,  \Verb+Name_i nin Known+ means $Name\_i \notin Known$.
\item \Verb+un(Known,{Name_i},Known_)+ means $Known\_ = Known \cup \{Name\_i\}$.
\item The last statement implements the state change. 
\end{itemize}

Let's now see the definition of \texttt{nameAlreadyExists}:
\begin{Verbatim}
nameAlreadyExists(Known,Birthday,Name_i,Known_,Birthday_) :-
  Name_i in Known &
  Known_ = Known &
  Birthday_ = Birthday.
\end{Verbatim}
where we can see how to say that there's no state change.

The order in which clauses are given is irrelevant in the sense that a clause can be defined before or after of being invoked for the first time.

The next schema to be translated is $FindBirthday$.
\begin{Verbatim}
findBirthday(Known,Birthday,Name_i,Date_o,Known_,Birthday_) :-
  findBirthdayOk(Known,Birthday,Name_i,Date_o,Known_,Birthday_)
  or
  notAFriend(Known,Birthday,Name_i,Known_,Birthday_).
\end{Verbatim}
where we can see that the translation starts to be repetitive. Now we show the definition of \Verb+findBirthdayOk+ and \Verb+notAFriend+:
\begin{Verbatim}
findBirthdayOk(Known,Birthday,Name_i,Date_o,Known_,Birthday_) :-
  Name_i in Known & 
  apply(Birthday,Name_i,Date_o) &
  Known_ = Known &
  Birthday_ = Birthday.

notAFriend(Known,Birthday,Name_i,Known_,Birthday_) :-
  Name_i nin Known & 
  Known_ = Known &
  Birthday_ = Birthday.
\end{Verbatim}
where we can see how to use output variables. We can also see the  \setlog \Verb+apply+ predicate which implements function application. That is, \Verb+apply(F,X,Y)+ is true if and only if $F(X) = Y$ holds. Note that \Verb+apply(F,X,Y)+ makes sense only if \Verb+F+ is a function and not any set. \setlog enforces this because, internally, \Verb+apply+ ``types'' \Verb+F+ with the predicate \Verb+pfun(F)+. Hence, if \Verb+apply+ is called with a first argument that happens not to be a function, the predicate will fail making \setlog to answer \Verb+no+.

Finally the translation of $Remind$ is the following:
\begin{Verbatim}
remind(Known,Birthday,Today_i,Cards_o,Known_,Birthday_) :-
  rres(Birthday,{Today_i},M) &
  dom(M,Cards_o) &
  Known_ = Known &
  Birthday_ = Birthday.
\end{Verbatim}
This is an interesting example because it shows how set and relational expressions must be translated. Given that in \setlog set and relational operators are implemented as predicates, it's impossible to write set and relational expressions. Instead, we have to introduce new variables (such as \Verb+M+) to ``chain'' the predicates. Predicate \Verb+rres(R,A,S)+ stands for $S = R \rres A$ while \Verb+dom(R,B)+ stands for $\dom R = B$ (that is, \Verb+dom(R,B)+ allows to compute the domain of \Verb+R+ and ``save'' it in \Verb+B+).

The schema specifying the initial state is translated as a clause receiving the state variables:
\begin{Verbatim}
birthdayBookInit(Known,Birthday) :-
  Known = {} &
  Birthday = {}.  
\end{Verbatim}
where \Verb+{}+ denotes the empty set ($\emptyset$).

The translation of $BirthdayBookInv$ is given in Section \ref{pruebas}.

\subsection{Types in \setlog}
So far we haven't given the types of the variables. \setlog is essentially an untyped formalism but in the last version we have added a type system similar to Z's. \setlog's type system is described in detail in chapter 9 of \setlog user's manual. Here we will give a broad description of how to use types in \setlog.

\setlog's type system allows users to define type synonyms to simplify the type declaration of clauses and variables. For example, we can define the following type synonyms:
\begin{Verbatim}
:- dec_type(bb,stype([name,date])).
:- dec_type(kn,stype(name)).
\end{Verbatim}
where \Verb+bb+ is a type identifier o synonym of the type \Verb+stype([name,date])+. In \Verb+stype([name,date])+, \Verb+name+ and \Verb+date+ correspond to the basic types $NAME$ and $DATE$ of the Z specification. In \setlog, Z basic types (called \emph{uninterpreted types} in \setlog) can be introduced without any previous declaration. In \setlog basic types must begin with a lowercase letter (i.e. they are constants). Besides, \Verb+[name,date]+ corresponds to the Cartesian product between \Verb+name+ and \Verb+date+ (i.e., is equivalent to $NAME \cross DATE$ in Z). In turn, \Verb+stype([name,date])+ corresponds to the type of all the sets of type \Verb+[name,date]+ (i.e., is equivalent to $\power(NAME \cross DATE)$ in Z). Hence, \Verb+stype([name,date])+ corresponds to the type of all the binary relations between \Verb+name+ and \Verb+date+ (i.e., $NAME \rel DATE$ in Z).

These type synonyms allow us to declare the type of the  \Verb+addBirthdayOk+ clause:
\begin{Verbatim}
:- dec_p_type(addBirthdayOk(kn,bb,name,date,kn,bb)).
\end{Verbatim}
This declaration must come before the clause definition:
\begin{Verbatim}
:- dec_p_type(addBirthdayOk(kn,bb,name,date,kn,bb)).
addBirthdayOk(Known,Birthday,Name_i,Date_i,Known_,Birthday_) :-
  Name_i nin Known &
  ...
\end{Verbatim}

The \Verb+dec_p_type+ predicate has only one argument of the following form:
\begin{Verbatim}
clause_name(parameters)
\end{Verbatim}
In turn, \Verb+parameters+ is a list whose elements corresponds one-to-one to the clause arguments. In this way, the type of each clause argument is given. Then, the type of \Verb+Known+ is \Verb+kn+, the type of \Verb+Birthday+ is \Verb+bb+, etc. 

\vspace{3mm}
\colorbox{gray!20}{%
\parbox{.9\textwidth}{%
In this version of \setlog clauses must be defined before they are used, due to the introduction of types. We remark this because this wasn't the case in previous versions of \setlog and isn't the case in Prolog programs. This means that we should first define \Verb+addBirthdayOk+ or \Verb+nameAlreadyExists+ and then \Verb+addBirthday+ because the latter invokes the former two.
}}
\vspace{3mm}

Note that above we didn't respect this rule because we decided to introduce types later for pedagogy reasons. From now on we will proceed according to the above rule; that is, by defining the clauses before they are used.

The following is the typed version of the \Verb+remid+ clause.
\begin{Verbatim}
:- dec_p_type(remind(kn,bb,date,kn,kn,bb)).
remind(Known,Birthday,Today_i,Cards_o,Known_,Birthday_) :-
  rres(Birthday,{Today_i},M) & dec(M,bb) &
  dom(M,Cards_o) &
  Known_ = Known &
  Birthday_ = Birthday.
\end{Verbatim}
This clause is interesting because it shows how clause variables are typed by means of the \Verb+dec(V,t)+ predicate. Indeed, \Verb+dec(V,t)+ is interpreted as ``variable \Verb+V+ is of tye \Verb+t+''.

The \setlog code including type declarations of the complete translation of the birthday book can be found in Appendix \ref{ap:bb}.

Recall that partial functions \emph{aren't} a type in Z. The same happens in \setlog; in fact it is impossible to define the type of all partial functions. The natural numbers are another example of a set that isn't a type. This means that if in Z we have $f: X \pfun Y$ in \setlog we declare \Verb+F+ to be of type \Verb+stype([x,y])+ and then we should either establish or prove that \Verb+F+ is a function (we'll see further details on this later). Likewise, if in Z we declare $x:\nat$ in \setlog we must declare \Verb+X+ to be of type \Verb+int+ and then either establish or prove that \Verb+0 =< X+ holds. In general, when a Z specification is translated into \setlog it would be convenient to first normalize the Z specification and then start the translation into \setlog. In this case the Z types are translated straightforwardly and the predicates introduced due to the normalization process become constraints at the \setlog level (i.e. \Verb+0 =< X+) or they are proved to be invariants. For instance, $x:\nat$ is a non-normalized declaration because $\nat$ isn't a type (it's a set). The normalized declaration would be $x:\num$ plus $x \geq 0$ conjoined in the predicate part. In this case, in \setlog the type of $x$ is \Verb+int+ and we should either establish or prove that $x$ is always greater than or equal to zero. 

In the birthday book the declaration $birthday:NAME \pfun DATE$ isn't normalized. The normalized declaration would be $birthday:NAME \rel DATE$ plus $birthday \in NAME \pfun DATE$ in the predicate part. In this case, $birthday$'s type is translated as we did above and we could add \Verb+pfun(Birthday)+ to the clauses where we use \Verb+Birthday+. \Verb+pfun+ is a \setlog predicate implementing the definition of (partial) function as a subclass of the sets of ordered pairs. But then, why we haven't state \Verb+pfun(B)+ somewhere in \Verb+addBirthday+? Because the idea is to show how \setlog can be used to prove that \Verb+pfun(B)+ is a state invariant of the specification. In fact had we written the state schema $BirthdayBook$ with normalized declarations, it would have resulted as follows:
\begin{schema}{BirthdayBook}
known: \power NAME \\
birthday: NAME \rel DATE
\where
birthday \in NAME \pfun DATE
\end{schema}
where $birthday \in NAME \pfun DATE$ would be a state invariant by definition (recall Section 6 of ``Introduction to the Z notation''). Therefore, we should have written the schema as follows:
\begin{schema}{BirthdayBook}
known: \power NAME \\
birthday: NAME \rel DATE
\end{schema}
\begin{schema}{BirthdayBookInv}
BirthdayBook
\where
known = \dom birthday \\
birthday \in NAME \pfun DATE
\end{schema}
where it's clear that we must \emph{prove} that \Verb+pfun(B)+ is a state invariant of the specification. We'll come back to this point in Section \ref{pruebas}.

\subsection{An alternative to encoding state variables}
The state schema of the birthday book specification has two variables ($known$ and $birthday$). These variables are included as parameters in the clauses encoding the operations of the specification; besides, the after state variables must also be included (those ending with `\Verb+_+'). This can become complex if the Z specification declares many state variables. In this case one should define \setlog clauses with many parameters which would make the code hard to read.

State variables can also be encoded by bundling them in a Prolog list. Later Prolog unification is used to access each state variable. The following \setlog code uses this encoding in \Verb+addBirthdayOk+\footnote{Here we don't include type declarations to simplify the presentation.}:
\begin{Verbatim}
addBirthdayOk(BirthdayBook,Name_i,Date_i,BirthdayBook_) :-
  BirthdayBook = [Known,Birthday] &
  Name_i nin Known &
  un(Known,{Name_i},Known_) &
  un(Birthday,{[Name_i,Date_i]},Birthday_) &
  BirthdayBook_ = [Known_,Birthday_].
\end{Verbatim}
where \Verb+BirthdayBook+ represents the before state and  \Verb+BirthdayBook_+ the after state. Both variables are assumed to be lists of length two. This assumption is enforced by means of the equalities:
\begin{Verbatim}
BirthdayBook = [Known,Birthday]
BirthdayBook_ = [Known_,Birthday_]
\end{Verbatim}
This means that if for any reason \Verb+addBirthdayOk+ is invoked with a first parameter that doesn't unify with \Verb+[Known,Birthday]+, the invocation will fail (same with the second parameter). 

Observe how the unification between \Verb+BirthdayBook_+ with the list \Verb+[Known_,+ \Verb+Birthday_]+ sets the after state. See that the list contains the same variables used in the postcondition of the operation.

It is possible to unify \Verb+BirthdayBook+ with a list containing any two variables; they need not to be the same used in the specification. For example, \Verb+BirthdayBook = [K,B]+.

Note that the first and last arguments of \Verb+addBirthdayOk+ are named as the state schema of the Z specification (i.e. $BirthdayBook$). This convention helps to relate the Z specification with the \setlog code, but it is in no way mandatory.

Clearly, encoding state variables in this way carries the complexity of a large number of arguments to the length of the list representing the state of the system. Hence, it is at the programmer discretion what is the best encoding for each situation. In fact, both encodings can coexist. In this case care must be taken when invoking clauses.

A slight advantage of using the encoding with Prolog lists can be seen in operations not using some of the state variables. For instance, the Z operation named $Remind$ doesn't use state variable $known$. In that case it can be encoded as follows:
\begin{Verbatim}
remind(BirthdayBook,Today_i,Cards_o,BirthdayBook_) :-
  BirthdayBook = [_,B] &
  rres(B,{Today_i},M) &
  dom(M,Cards_o) &
  BirthdayBook_ = BirthdayBook.
\end{Verbatim}
where we have written `\Verb+_+' in place of the first element of the list unifying with \Verb+BirthdayBook+. This means that this element of the list doesn't matter; it can be anything. If the list of state variables has many elements and some of them are not used in a clause, the programmer can use `\Verb+_+' in different positions instead of writing variable names.

Another benefit of using the list-based encoding is when stating that no state change occurs. For example, in the birthday book specification \Verb+BirthdayBook_ = BirthdayBook+ is written instead of \Verb+Known_ = Known & Birthday_ = Birthday+. This encoding is a good help when there are many state variables.

If the list of state variables is long it might be a problem recalling what is the position of each variable. Switching the order of two or more variables will lead to unsound programs.

\subsection{Translating ordered pairs}
Ordered pairs are encoded as Prolog lists of two elements. For instance, if $x$ is a variable $(x,3)$ is translated as \Verb+[X,3]+.

The translation of $p: \num \cross V$ and $p$.$1 = x - 4$ is: \Verb+P = [A,_] & A is X - 4+ (see the \Verb+is+ operator in Section \ref{aritmetica}). This means that, basically, we use unification to force that \verb+P+ be a two elements list such that its first element is variable \Verb+A+, which in turn must be equal to \Verb+X - 4+. \Verb+A+ must be a variable name not used in the clause. Note that we write `\Verb+_+' as the second argument because we aren't interested in the second component of \Verb+P+.

Prolog lists must not be used to encode Z sequences.

\subsection{Translating sets}

\subsubsection{Extensional sets --- Introduction to set unification}
The set $\{1,2,3\}$ is simply translated as \Verb+{1,2,3}+. If one of the elements of the set is a variable or an element of an enumerated type, take care of the differences concerning variables and constants in Z and \setlog. For example, if in Z $x$ is a variable, then the set $\{2,x,6\}$ is translated as \Verb+{2,X,6}+.

However, \setlog provides a form of extensional sets that, in a sense, is more powerful than the one offered by Z. The term \Verb+{.../...}+ is called \emph{extensional set constructor}. In \Verb+{E/C}+ the second argument (i.e. \Verb+C+) must be a set. \Verb+{E/C}+ means $\{E\} \cup C$. Then, there are solutions where $E \in C$. To avoid such solutions (in case they're incorrect or unwanted) the predicate $E \notin C$ must be explicitly added to the formula. In order to make the language more simple, \setlog accepts and prints terms such as  \Verb+{1,2 / X}+ instead of \Verb+{1 / {2 / X}}+.

The extensional set constructor is useful and in general it's more efficient than other encodings. For example, the Z predicate:
\[
A' = A \setminus \{d?\}
\]
can be translated by means of the \setlog predicate \Verb+diff+, whose semantics is equivalent to $\setminus$ (see Table \ref{t:setoper}):
\begin{Verbatim}
diff(A,{D_i},A_)
\end{Verbatim}
Bu it also can be translated by means of an extensional set:
\begin{Verbatim}
A = {D_i / A_} & D_i nin A_ or D_i nin A & A_ = A
\end{Verbatim}
which in general is more efficient.

That is, the predicate \Verb+A = {D_i / A_}+ \emph{unifies} \Verb+A+ with \Verb+{D_i / A_}+ in such a way that it finds values for the variables to make the equality true. If such values don't exist the unification fails and \setlog tries the second disjunct. 

Why we conjoined \Verb+D_i nin A_+? Simply because, for instance, \Verb+A = {1,2}+, \Verb+D_i = 1+ and \Verb+A_ = {1,2}+ is a solution of the equation but it isn't a solution of $A' = A \setminus \{d?\}$. Precisely, when \Verb+D_i nin A_+ is conjoined all the solutions where \Verb+D_i+ belongs to \Verb+A_+ are eliminated.

\setlog solves equalities of the form \Verb+B = C+, where \Verb+B+ and \Verb+C+ are terms denoting sets, by using \emph{set unification}. Se unification is at the base of the deductive power of \setlog making it an important extension of Prolog's unification algorithm. Set unification is inherently computationally hard because finding out whether or not two sets are equal implies, in the worst case, computing all the permutations of their elements. On top of that, it is the fact that \setlog can deal with \emph{partially specified} sets, that is sets where some of their elements or part of the set are variables. For these reasons, in general, \setlog will show efficiency problems when dealing with certain formulas but, at the same time, we aren't aware of other tools capable of solving some of the problems \setlog can.

\subsubsection{Cartesian products}
In \setlog Cartesian products are written \Verb+cp(A,B)+ where \Verb+A+ and \Verb+B+ can be variables, extensional sets and Cartesian products.

\subsubsection{Integer intervals}
A Z integer interval such as $n \upto m$ is translated as \Verb+int(n,m)+.

\subsection{Translating function application}
One interesting application of set unification is the application of a function to its argument. Given that partial functions are frequently used in Z it's necessary to add predicates of the form $x \in \dom f$, before attempting to apply $f$ to $x$. The translation of these formulas into \setlog can be done by using the predicate \Verb+apply+ or by using a set membership predicate which leads to set unification. For example the Z formula:
\begin{zed}
x \in \dom f \land f(x) = y
\end{zed}
can be translated in a direct fashion:
\begin{Verbatim}
dom(F,D) & X in D & apply(F,X,Y)
\end{Verbatim}
or using a set membership predicate:
\begin{Verbatim}
[X,Y] in F & pfun(F)
\end{Verbatim}
which is equivalent and more efficient as it doesn't require to  ``compute'' the domain of \Verb+F+. Concerning the equivalence of both formulas, note that \Verb+[X,Y] in F+ is transformed into \Verb+F = {[X,Y] / G}+ (for some \Verb+G+). Then, if there are no ordered pairs in \Verb+F+ whose first component is \Verb+X+, the unification will fail which is equivalent to say that $x \in \dom f$ is false. In the same way, if there's an ordered pair in \Verb+F+ whose first component is \Verb+X+ but whose second component is not \Verb+Y+, the unification will fail as well (which is equivalent to $f(x) \neq y$). Variable \Verb+G+ is automatically generated by \setlog in such a way that it will be new in the formula. This is interpreted as ``there exists \Verb+G+ such that\dots''.

However, any of the encodings using \Verb+pfun+ might not be correct. In effect, the fact that \Verb+F+ is a function might be something that should be proved to hold, usually in the form of a state invariant (recall schema $BirthdayBookInv$ shown above and the discussion around it). Consequently, if we use \Verb+apply+ or \Verb+pfun+ we would be establishing that \Verb+F+ is a function rather than getting it as a consequence of the encoding. In other words, if we write \Verb+pfun(F)+ we would be indicating to the programmer to control that the data structure used to encode \Verb+F+ to be a function every time that function is applied to an argument. Usually, the most reasonable implementation is for the code to guarantee that the data structure is a function without explicitly controlling that.

Therefore, we can't include \Verb+apply(F,...)+ nor \Verb+pfun(F)+ in the operations of the model if \Verb+pfun(F)+ is an invariant to be proved because otherwise we would be cheeting. Then, the question is, how $f~x = y$ is translated when we only know that $f$ is a binary relation and that $x \in \dom f$? The \setlog code is the following:
\begin{Verbatim}
F = {[X,Y] / G} & [X,Y] nin G & comp({[X,X]},G,{})
\end{Verbatim}
The justification is as follows. If we know that $x \in \dom f$ the there exist \Verb+Y+ and \Verb+G+ such that \Verb+F = {[X,Y] /+ \Verb+G} & [X,Y] nin G+, due to the above analysis. Besides, if we are saying that we can apply $f$ to $x$ is because there is one and only one ordered pair in $f$ whose first component is $x$. Note that we aren't saying that $f$ is a function, we're just saying that $f$ is \emph{locally} a function in $x$ (it might well be a function in other points of its domain but we don't know that yet). Saying that in $f$ there is exactly one ordered pair whose first component is $x$ is the same than saying that there are no ordered pairs in \Verb+G+ whose first component is $x$. We say this by using the composition operator defined over binary relations, namely \Verb+comp+ (see Table \ref{t:reloper}), when we conjoin the predicate \Verb+comp({[X,X]},G,{})+. Indeed, this predicate says that when \Verb+{[X,X]}+ is composed with \Verb+G+ the result is the empty set. This can happen for two reasons: \Verb+G+ is the empty binary relation, in which case it's obvious that there are no ordered pairs with first component \Verb+X+; or \Verb+G+ is non-empty but no pair in it composes with \Verb+[X,X]+, which is equivalent to say that \Verb+X+ does not belong to the domain of \Verb+G+. We could have said the same by stating that \Verb+dom(G,D) & X nin D+ but this is usually less efficient because it requires to compute the domain of \Verb+G+.

The \setlog library \Verb+setloglibpf.slog+ defines the predicate \Verb+applyTo(F,X,Y)+ which implements the code shown above. This library can be loaded with \Verb+add_lib('setloglibpf.slog')+.

Observe that in \Verb+findBirthdayOk+ we have used \Verb+apply+ which, after the above analysis, is incorrect because  \Verb+pfun(Birthday)+ is intended to be an invariant of the program. We should replace \Verb+apply+ by \Verb+applyTo+. We didn't do it in that way because we think that it requires a rather complex explanation when we were just introducing \setlog.

\subsection{\label{aritmetica}Translating arithmetic expressions}
Almost all Z arithmetic expressions are translated directly into \setlog, with some exceptions. The relational symbols $\leq$, $\geq$ and $\neq$ are translated as \Verb+=<+, \Verb+>=+ and \Verb+neq+, respectively. The arithmetic operators are the usual ones: \Verb.+., \Verb.-., \Verb.*., \Verb.div. y \Verb.mod..  

An equality of the form $x' = x + 1$ is translated as \Verb!X_ is X + 1! (that is, in arithmetic equalities you mustn't use \Verb+=+ but \Verb+is+). Furthermore, if in Z we have $A = \{x, y - 4\}$ ($A$, $x$ and $y$ variables) it has to be encoded as: \Verb+A = {X,Z} & Z is Y - 4+, where \Verb+Z+ is a variable not used in the clause. The problem is that nor \setlog nor Prolog evaluate arithmetic expression unless the programmer forces it by using the \Verb+is+ operator. This means that if in \setlog we run \Verb+{X,Y - 4} = {Y - 3 - 1,X}+, the answer will be \Verb+no+ because \setlog will try to find out whether or not \Verb+Y - 4 = Y - 3 - 1+ without evaluating the expressions (that is, it will consider them, basically, as character strings where \Verb+Y+ is an integer variable and thus it is impossible for the equality to hold regardless of the value of \Verb+Y+). On the contrary, if we run \Verb+{X,A} = {B,X} & A is Y - 4 & B is Y - 3 - 1+ \setlog will return several solutions (with some repetitions), meaning that the sets are equal in several ways.

The same applies to the \Verb+neq+ predicate: for \setlog \Verb+Y - 4 neq Y - 3 - 1+ is true. As a consequence we must write: \Verb+H is Y - 4 & U is Y - 3 - 1 & H neq U+. However, this is not necessary with the order predicates: \Verb.X + 1 > X. is satisfiable but \Verb+X - 1 > X+ isn't.

On the other hand, $A \subseteq \nat$ or $A:\power \nat$ are translated with a restricted universal quantifier (see Section \ref{quatifiers}):
\begin{Verbatim}
foreach(X in A, 0 =< X)
\end{Verbatim}
In any case, usually, $A \subseteq \nat$ and $A:\power \nat$ are invariants. If this is the intention, then they should be proved to be rather than establishing them.

\subsection{Translating set operators}
Set, relational, functional and sequence operators are translated as shown in Tables \ref{t:setoper}, \ref{t:reloper} and \ref{t:listoper}.

In order to be able to work with the sequence operators shown in Table \ref{t:listoper} load the corresponding library file (e.g. \Verb+consult('setlogliblist.slog')+) into the \setlog environment. 

\begin{table}
\begin{tabularx}{\textwidth}{Xll}
\hline\hline
\multicolumn{1}{c}{\textsc{Operator}} &
\multicolumn{1}{c}{\setlog} &
\multicolumn{1}{c}{\textsc{Meaning}} \\\hline
set           & \Verb+\set(A)+        & $A$ is a set \\
equality             & \Verb+A = B+          & $A = B$ \\
set membership          & \Verb+x \In A+        & $x \in A$ \\
union                & \Verb+\Cup(A,B,C)+    & $C = A \cup B$ \\
intersection         & \Verb+\Cap(A,B,C)+    & $C = A \cap B$ \\
difference         & \Verb+\Diff(A,B,C)+   & $C = A \setminus B$ \\
subset          & \Verb+\Subseteq(A,B)+ & $A \subseteq B$ \\
strict subset & \Verb+\Subset(A,B)+   & $A \subset B$ \\
disjointness  & \Verb+\Disj(A,B) +    & $A \disj B$ \\
cardinality         & \Verb+\Size(A,n)+     & $\card{A} = n$ \\
\hline
\multicolumn{3}{c}{\textsc{Negations}} \\\hline
%
equality      & \Verb+A \Neq B+        & $A \neq B$ \\
set membership    & \Verb+x \Nin A+        & $x \notin A$ \\
union         & \Verb+\Ncup(A,B,C)+    & $C \neq A \cup B$ \\
intersection  & \Verb+\Ncap(A,B,C)+    & $C \neq A \cap B$ \\
difference    & \Verb+\Ndiff(A,B,C)+   & $C \neq A \setminus B$ \\
subset       & \Verb+\Nsubseteq(A,B)+ & $A \not\subseteq B$ \\
disjointness & \Verb+\Ndisj(A,B)+    & $A \not\disj B$ \\
\hline\hline
\end{tabularx}
\caption{\label{t:setoper} Set operators available in \setlog}
\end{table}

\begin{table}
\begin{tabularx}{\textwidth}{Xll}
\hline\hline
\multicolumn{1}{c}{\textsc{Operator}} &
\multicolumn{1}{c}{\setlog} &
\multicolumn{1}{c}{\textsc{Meaning}} \\\hline
binary relation        & \Verb+\Rel(R)+       & $R$ is a binary relation \\
partial function         & \Verb+\Pfun(R)+ & $R$ is a partial function \\
function application   & \Verb+\Apply(f,x,y)+ & $f(x) = y$ \\
domain                 & \Verb+\Dom(R,A)+     & $\dom R = A$ \\
range                   & \Verb+\Ran(R,A)+     & $\ran R = A$ \\
composition             & \Verb+\Comp(R,S,T)+  & $T = R \circ S$ \\
inverse                 & \Verb+\Inv(R,S)+     & $S = R^{-1}$ \\
domain restriction  & \Verb+\Dres(A,R,S)+  & $S = A \dres R$ \\
domain anti-restriction & \Verb+\Ndres(A,R,S)+ & $S = A \ndres R$ \\
range restriction  & \Verb+\Rres(A,R,S)+  & $S = R \rres A$ \\
range anti-restriction  & \Verb+\Nrres(A,R,S)+ & $S = R \nrres A$ \\
update          & \Verb+\Oplus(R,S,T)+ & $T = R \oplus S$ \\
relational image      & \Verb+\Rimg(R,A,B)+  & $B = R[A]$ \\\hline
\multicolumn{3}{c}{\textsc{Negaciones}} \\\hline
\multicolumn{3}{l}{\parbox{.95\textwidth}{\vspace{2pt}
All negations are written by prefixing a letter \Verb+n+ to the corresponding operator. For example, the negation of \Verb+\Dom(R,A)+ is \Verb+\Ndom(R,A)+, that of \Verb+\Ndres(A,R,S)+ is \Verb+n\Ndres(A,R,S)+, etc.}}
  \\[7pt]\hline\hline
\end{tabularx}
\caption{\label{t:reloper} Relational operators available in \setlog}
\end{table}

\begin{table}
\begin{tabularx}{\textwidth}{Xll}
\hline\hline
\multicolumn{1}{c}{\textsc{Operator}} &
\multicolumn{1}{c}{\setlog} &
\multicolumn{1}{c}{\textsc{Meaning}} \\\hline
sequence         & \Verb+\List(s)+        & $s$ is a sequence \\
extensional sequence & \verb+{[1,a],[2,b],...,[n,z]}+ & $\langle a,b,\dots,z\rangle$ \\
head       & \Verb+\Head(s,e)+      & $e = head~s$ \\
tail           & \Verb+\Tail(s,t)+      & $t = tail~s$ \\
last         & \Verb+\Last(s,e)+      & $e = last~s$ \\
front         & \Verb+\Front(s,t)+     & $t = front~s$ \\
add (cons) & \Verb+\Add(s,e,t)+     & $t = s \cat \langle e \rangle$ \\
concatenation  & \Verb+\Concat(s,t,u)+  & $u = s \cat t$ \\
filter       & \Verb+\Filter(A,s,t)+  & $t = A \filter s$ \\
extraction     & \Verb+\Extract(s,A,t)+ & $t = s \extract A$ \\
\hline\hline
\end{tabularx}
\caption{\label{t:listoper} Sequence operators available in \setlog}
\end{table}

The cardinality operator accepts as second argument only a constant or a variable. Hence, if we run \Verb!size(A,X + 1)! \setlog answers \Verb+no+; instead if we run \Verb!size(A,Y) &! \Verb!Y is X + 1! (\Verb+Y+ must be a variable not used in the clause) the answer is \Verb+true+ because the formula is satisfiable. \setlog will answer \Verb+no+ if we execute \Verb!size(A,Y) & Y = X + 1!.

\subsection{\label{logicos}Translating logical operators}
Only logical conjunction (\Verb+&+) and disjunction (\Verb+or+) are available in \setlog. Logical negation ($\lnot$) doesn't exist because it's replaced by the negations of the operators available in \setlog. For instance, if we want to translate $\lnot x \in A$ we write in \setlog \Verb+X nin A+. In the same way, $\lnot A = b$ is translated as \Verb+A neq b+. In general, for each set, relational and arithmetic operator there exists a \setlog predicate implementing its negation. For instance, the Z predicate $A \not\subseteq B$ is translated as \Verb+nsubset(A,B)+; and $\lnot a \leq y$ as \Verb+A > Y+. Tables \ref{t:setoper} and \ref{t:reloper} include the negation for every set theoretic operator.

As a summary, a Z schema like the following one:
\begin{schema}{WithdrawE}
\Xi Bank \\
n?:NIC \\
a?: MONEY \\
msg!: MSG
\where
\lnot(n? \in \dom sa \land a? \leq sa(n?) \land a? > 0) \\
msg! = error
\end{schema}
can be translated as follows (we assume that schema $Bank$ only declares variable $sa$):
\begin{Verbatim}
withdrawE(Sa,N_i,A_i,Msg_o,Sa_) :-
  (dom(Sa,D) &
   N_i nin D
   or
   apply(Sa,N_i,Y) &
   A_i > Y
   or
   A_i =< 0
  ) &
  Msg_o = error &
  Sa_ = Sa.
\end{Verbatim}

The remaining propositional connectives can be encoded by means of the well-known propositional equivalencies in terms of $\land$, $\lor$ and $\lnot$. For example, $p \implies q$ is first written as $\lnot p \lor q$ and then $\lnot p$ and $q$ are translated into \setlog.

\subsubsection{\label{quatifiers}Quantifiers}
In general existential quantifiers need not to be translated because \setlog semantics is based on existentially quantifying all variables of any given program. For example, if in Z we have:
\[
\exists x:\nat | x \in A
\]
it can be translated as:
\begin{Verbatim}
0 =< X & X in A
\end{Verbatim}
because the semantics of the \setlog program is, essentially, an existential quantifier over both variables.

Things are different when dealing with universal quantifiers. In \setlog we only have so-called \emph{restricted universal quantifiers} (RUQ). A RUQ is a formula of the following form:
\[
\forall x \in A : P(x)
\]
where $P$ is a proposition depending on $x$. In \setlog the simplest RUQ are encoded as follows:
\begin{Verbatim}
foreach(X in A,P(X))
\end{Verbatim}
There are more complex and expressive RUQ available\footnote{Have a look at chapter 6 of \setlog user's manual.}.

Recall that a proper use of the Z language avoids most of the quantified formulas.

\subsection{\label{da}Translating axiomatic definitions}
There's no simple way of translating all axiomatic definitions into \setlog because they are so general and serve to so many purposes. For this reason we will show how to translate the most used forms of axiomatic definitions.

One of the problems when translating axiomatic definitions is that they are global objects while in \setlog global objects don't really exist. 

The easiest and simplest way of translating axiomatic definitions is defining a \setlog clause including all the predicates of the axiomatic definition. Here we're assuming that each axiomatic definition declares only one variable. As always, all the restrictions on constant and variable names as well as the considerations on translating types, must be carefully watched.

Let's see the most common cases.

\bigskip

\begin{samepage}
\hrule\vspace{1mm}\noindent
Z specification.
\begin{axdef}
root:USR
\end{axdef}
\setlog code.
\begin{Verbatim}
:- dec_p_type(root_da(usr)).
root_ad(X) :- X = root.
\end{Verbatim}
\hrule
\end{samepage}

\bigskip

\begin{samepage}
\hrule\vspace{1mm}\noindent
Z specification.
\begin{axdef}
adm:\power USR
\where
adm = \{root,sec\}
\end{axdef}
\setlog code.
\begin{Verbatim}
:- dec_p_type(adm_da(stype(usr))).
adm_ad(X) :- X = {root,sec}.
\end{Verbatim}
\hrule
\end{samepage}

\bigskip

As shown in the following example, the translation of some axiomatic definitions uses \setlog recursive definitions.

\begin{samepage}
\vspace{2mm}
\hrule\vspace{1mm}\noindent
Z specification.
\begin{axdef}
sum:\seq \num \fun \num
\where
sum~\langle\rangle = 0 \\
\forall s:\seq \num; n:\num \bullet sum(s \cat \langle n \rangle) = n + sum~s
\end{axdef}
\setlog code. Recall that lists are sets of ordered pairs.
\begin{Verbatim}
:- dec_p_type(sum_da(stype([int,int]),int))).
sum_ad({},0).
sum_ad({[N,X]/S},Sum) :- 
  [N,X] nin S & Sum is X + Sum1 & sum_ad(S,Sum1).
\end{Verbatim}
Note that the type of \Verb+sum_da+ is declared only once.
\hrule
\end{samepage}

\bigskip

These clauses can be invoked from clauses implementing schemas or other axiomatic definitions. For example, the following is part of the translation of an operation specifying how user $a$ creates the system account of user $u$:
\begin{Verbatim}
createUsr(...,A,U,...) :- ... & root_ad(A) & ...
\end{Verbatim}
which means that we're checking that user \Verb+A+ is $root$. The following version checks that \Verb+A+ is an administrator:
\begin{Verbatim}
createUsr(...,A,U,...) :- ... & adm_ad(Adm) & A in Adm...
\end{Verbatim}
where \Verb+Adm+ is a new variable. In these cases variable \Verb+A+ \emph{unifies} with the argument named \Verb+X+ in \Verb+root_ad+ and \Verb+adm_ad+. In the first case this implies that \Verb+A+ is equal to \Verb+root+. In the second case this implies that \Verb+A+ is equal to \Verb+{root,sec}+.

\section{\label{simulacion}Simulating \setlog prototypes}
A \setlog implementation of a Z specification is easy to get but usually it won't meet the typical performance requirements demanded by users. Hence, we see a \setlog implementation of a Z specification more as a \emph{prototype} than as a final program. On the other hand, given the similarities between a Z specification and the corresponding \setlog program, it's reasonable to think that the prototype is a \emph{correct} implementation of the specification\footnote{In fact, the translation process can be automated in many cases.}. Then, we can use these prototypes to make an early validation of the requirements. 

Validating user requirements by means of prototypes entails executing the prototypes together with the users so they can agree or disagree with the behavior of the prototypes. This early validation will detect many errors, ambiguities and incompleteness present in the requirements and possible misunderstandings or misinterpretations generated by the software engineers. Without this validation many of these issues would be detected in later stages of the project thus increasing the project costs. Think that if one of these issues is detected once the product has been delivered it means to correct it from the requirements document, the specification, the design, the implementation, the user documentation, etc.

Since we see \setlog programs as prototypes we talk about \emph{simulations} or \emph{animations} rather than \emph{executions} when speaking about running them. However, technically, what we do is no more than running a program. The word \emph{simulation} is usually used in the context of \emph{models} (e.g. modeling and simulation). In a sense, our \setlog programs are \emph{executable models} of the user requirements. On the other hand, the word \emph{animation} is usually used in the context of formal specifications. In this sense, the \setlog implementation of a Z specification can be seen as an \emph{executable specification}. In fact, as we will see, \setlog programs have features and properties usually enjoyed by specifications and models, which are rare or nonexistent in programs written in imperative (and even functional) programming languages.

Be it execution, simulation or animation the basic idea is to provide inputs to the program, model or specification and observe the produced outputs or effects. Besides, we will show that \setlog offers more possibilities beyond this basic idea.

\subsection{\label{simples}Basic simulations}
Let's see an example of a simulation on a \setlog prototype. Assume the prototype of the birthday book is saved in a file named \Verb+bb.slog+. We start by executing the Prolog interpreter from a   command terminal and from the folder where \setlog was installed\footnote{The name of the Prolog executable may vary depending on the interpreter and the operating system. The example corresponds to a Ubuntu Linux machine and SWI-Prolog.}.
\begin{verbatim}
~/setlog$ prolog

?- consult('setlog.slog').

?- setlog.

{log}=> consult('bb.slog').

{log}=> birthdayBookInit(K,B) & addBirthday(K,B,maxi,160367,K_,B_).
K = {},  
B = {},  
K_ = {maxi},  
B_ = {[maxi,160367]}

Another solution?  (y/n) y
no
{log}=> 
\end{verbatim}
where in each line we're doing the following:
\begin{enumerate}
\item The Prolog interpreter is executed.
\item The \setlog interpreter is loaded.
\item The \setlog interpreter is accessed.
\item The birthday book prototype is loaded.
\item The simulation is run:
\begin{Verbatim}
birthdayBookInit(K,B) & addBirthday(K,B,maxi,160367,K_,B_).
\end{Verbatim}
consisting of:
\begin{itemize}
\item \Verb+birthdayBookInit+ is called passing to it any two variables as arguments;
\item \Verb+addBirthday+ is called passing to it in the first and second arguments the same variables used to call \Verb+birthdayBookInit+; as the third and fourth arguments two constants; and two new variables in the last two parameters.
\end{itemize}
Observe that the simulation ends in a dot.
\item \setlog shows the result of the simulation.
\item \setlog asks if we want to see other solutions and we answer yes.
\item \setlog says there are no more solutions.
\end{enumerate}

Let's see the simulation in detail:
\begin{Verbatim}
birthdayBookInit(K,B) & addBirthday(K,B,maxi,160367,K_,B_).
\end{Verbatim}
When we call \Verb+birthdayBookInit(K,B)+, \Verb+K+ and \Verb+B+ unify with  \Verb+Known+ and \Verb+Birthday+ which are the formal parameters used in the definition of \Verb+birthdayBookInit+ (see the complete code in Appendix \ref{ap:bb}). This implies that \Verb+K+ is equal to \Verb+Known+ and \Verb+B+ is equal to \Verb+Birthday+ which in turn implies that \Verb+K+ and \Verb+B+ are equal to \Verb+{}+. This is exactly the first line of the answer returned by \setlog. Hence, when \Verb+addBirthday(K,B,maxi,160367,K_,B_)+ is called, it's like we were calling:
\begin{Verbatim}
addBirthday({},{},maxi,160367,K_,B_)
\end{Verbatim}
Calling \Verb+addBirthday+ implies the \emph{non-deterministic} invocation of \Verb+addBirthdayOk+ and \texttt{nameAlreadyExists}. That is, both clauses are invoked in an unspecified order. Let's assume that \Verb+addBirthdayOk+ is invoked first. In this case, unification goes as follows:
\begin{Verbatim}
Known = {}
Birthday = {}
Name_i = maxi
Date_i = 160367
K_ = Known_
B_ = Birthday_
\end{Verbatim}
Hence \Verb+addBirthdayOk+ is instantiated as follows:
\begin{Verbatim}
maxi nin {} &
un({},{maxi},K_) &
un({},{[maxi,160367]},B_)
\end{Verbatim}
which reduces to:
\begin{Verbatim}
K_ = {maxi} &
B_ = {[maxi,160367]}
\end{Verbatim}
which corresponds to the second line of the answer returned by \setlog.

When '\Verb+y+' is pressed \setlog invokes  \texttt{nameAlreadyExists} because it was the clause pending of invocation. Again, unification takes place and a new series of equations are produced:
\begin{Verbatim}
Known = {}
Birthday = {}
Name_i = maxi
K_ = Known
B_ = Birthday
\end{Verbatim}
which implies that \Verb+K+ unifies with \Verb+{}+. Then, \Verb+nameAlreadyExists+ is instantiated as follows:
\begin{Verbatim}
maxi in {}
\end{Verbatim}
As this predicate is obviously false, the invocation to \Verb+nameAlreadyExists+ fails and hence \setlog produces no solution. As a consequence \setlog answers \Verb+no+ after we press '\Verb+y+'.

The following simulation is longer and includes the previous one.
\begin{Verbatim}
birthdayBookInit(K,B)                  & addBirthday(K,B,maxi,160367,K1,B1) &
addBirthday(K1,B1,'Yo',201166,K2,B2)   & findBirthday(K2,B2,'Yo',C,K3,B3) &
addBirthday(K3,B3,'Otro',201166,K4,B4) & remind(K4,B4,160367,Card,K5,B5) &
remind(K5,B5,201166,Card1,K_,B_).
\end{Verbatim}
Here we can see that we're calling all the operations defined in the prototype; that we use different variables to chain the state transitions; and that it's possible to use constants beginning with an uppercase letter as long as we enclose them between single quotation marks.

The first solution returned by that simulation is the following:
\begin{Verbatim}
K = {},  
B = {},  
K1 = {maxi},  
B1 = {[maxi,160367]},  
K2 = {maxi,Yo},  
B2 = {[maxi,160367],[Yo,201166]},  
C = 201166,  
K3 = {maxi,Yo},  
B3 = {[maxi,160367],[Yo,201166]},  
K4 = {maxi,Yo,Otro},  
B4 = {[maxi,160367],[Yo,201166],[Otro,201166]},  
Card = {maxi},  
K5 = {maxi,Yo,Otro},  
B5 = {[maxi,160367],[Yo,201166],[Otro,201166]},  
Card1 = {Yo,Otro},  
K_ = {maxi,Yo,Otro},  
B_ = {[maxi,160367],[Yo,201166],[Otro,201166]}
\end{Verbatim}
where we can see that \setlog gives us the chance to have a complete trace of the prototype execution. Note also that \setlog eliminates the single quotation marks we used to enclose some constants.

It's important to remark that the variables used to chain the state transitions (i.e. \Verb+K1+, \Verb+B1+, \dots, \Verb+K5+, \Verb+B5+) must be all different. If done otherwise, the simulation might be incorrect. For instance:
\begin{Verbatim}
birthdayBookInit(K,B) & addBirthday(K,B,N,C,K,B).
\end{Verbatim}
will fail as the values of \Verb+K+ and \Verb+B+ before invoking \Verb+addBirthday+ can't unify with the values returned by it. In other words, the \Verb+K+ and \Verb+B+ as the first two arguments of \Verb+addBirthday+ can't have the same value than the \Verb+K+ and \Verb+B+ used as the last two arguments. We could use the same variable for the before and after state of query state operations (for instance when we invoke \Verb+findBirthday+ and \Verb+remid+).

So far the two simulations we have performed start in the initial state. It's quite simple to start a simulation from any state:
\begin{Verbatim}
K = {maxi,caro,cami,alvaro} &
B = {[maxi,160367],[caro,201166],[cami,290697],[alvaro,110400]} &
addBirthday(K,B,'Yo',160367,K1,B1) & remind(K1,B1,160367,Card,K1,B1).
\end{Verbatim}
where we can see that we use the same variable to indicate the before and after state of \Verb+remid+ (because we know this clause produces no state change). In this case the answer is:
\begin{Verbatim}
K = {maxi,caro,cami,alvaro},  
B = {[maxi,160367],[caro,201166],[cami,290697],[alvaro,110400]},  
K1 = {maxi,caro,cami,alvaro,Yo},  
B1 = {[maxi,160367],[caro,201166],[cami,290697],[alvaro,110400],[Yo,160367]},  
Card = {maxi,Yo}
\end{Verbatim}

A potential problem of manually defining the initial state for a simulation is that this state, due to human error, might not verify the state invariant. Nevertheless, it's very easy to avoid this problem as we will see in Section \ref{evaluacion}.

\subsection{Type checking and simulations}
So far we haven't really used \setlog's typechecker. Actually when we consulted \Verb+bb.slog+ the types weren't checked. In other words \setlog ignored the \Verb+dec_p_type+ assertions included in \Verb+bb.slog+. This means that possible type errors weren't detected by \setlog. In this sense \setlog executed all the simulations in untyped mode. In this section we'll see how to call the typechecker and how this affects simulations. Recall reading chapter 9 of \setlog user's manual for further details.

Type checking can be activated by means of the \Verb+type_check+ command which should be issued before the file is consulted.
\begin{verbatim}
~/setlog$ prolog

?- consult('setlog.slog').

?- setlog.

{log}=> type_check.         % typechecker is active

{log}=> consult('bb.slog').
\end{verbatim}
In this way, when \setlog executes command \Verb+consult+ it invokes the typechecker and if there are type errors we'll see an error message.

Type checking can be deactivated at any time by means of command \Verb+notype_check+.

When the typechecker is active all formulas (or programs or simulations) must be correctly typed because otherwise \setlog will just print a type error.
\begin{Verbatim}
{log}=> birthdayBookInit(K,B) & addBirthday(K,B,maxi,160367,K_,B_).

***ERROR***: type error: variable K has no type declaration
\end{Verbatim}
Then, we have to declare the type of all variables:
\begin{Verbatim}
{log}=> birthdayBookInit(K,B) & addBirthday(K,B,name?maxi,date?d160367,K_,B_) & 
        dec([K,K_],kn) & dec([B,B_],bb).

K = {},  
B = {},  
K_ = {name?maxi},  
B_ = {[name?maxi,date?d160367]}
\end{Verbatim}
Observe that in this case we have written \Verb+date?d160367+ instead of \Verb+date?160367+ (without `d') because in \setlog constants of a basic type \Verb+t+ are of the form \Verb+t?atom+ where \Verb+atom+ must be a Prolog atom (numbers aren't atoms).

If the user wants to typecheck the program, for instance \Verb+bb.slog+, but (s)he doesn't want to deal with types when running simulations, the typechecker can be deactivated right after consulting the program. In this way \setlog will check the types of the program but it then will accept untyped formulas (or programs or simulations).

Clearly, in general, working with untyped simulations is easier but more dangerous because we could call the program with ill-typed inputs thus causing unexpected behaviors.

In the rest of this section we'll work with untyped simulations. This means that the user must ensure that typechecking is deactivated (command \Verb+notype_check+).

\subsection{Simulaciones using integer numbers}
As we have said, \setlog is, essentially, a set solver. However, it's also capable of solving formulas containing predicates over the integer numbers. In that regard, \setlog uses two external solvers known as CLP(FD)\footnote{\url{https://www.swi-prolog.org/pldoc/man?section=clpfd-predicate-index}} and CLP(Q)\footnote{\url{https://www.swi-prolog.org/pldoc/man?section=clpqr}}. Each of them has its advantages and disadvantages.

By default \setlog uses CLP(Q). Users can change to CLP(FD) by means of command \Verb+int_solv+-\Verb+er(clpfd)+ and can come back to CLP(Q) by means of \Verb+int_solver(clpq)+.

Generally speaking, it's more convenient to run simulations when  CLP(FD) is active because it tends to generate more concrete solutions. In particular CLP(FD) is capable of performing labeling over the integer numbers which allows users to go through the solutions interactively. Labeling works if at least some of the integer variables are bound to a finite domain. Variable \Verb+N+ is bound to the finite domain \Verb+int(a,b)+ (\verb+a+ and \Verb+b+ integer numbers) if predicate \Verb+N in int(a,b)+ is in the formula. See chapter 7 of \setlog user's manual for more details.

For example, if CLP(Q) is active, the answer to the following goal:
\begin{verbatim}
Turn is 2*N + 1.
\end{verbatim}
is exactly the same formula. That is, \setlog is telling us that the formula is satisfiable but we don't have one of its solutions. If we activate CLP(FD):
\begin{verbatim}
int_solver(clpfd).

Turn is 2*N + 1.
\end{verbatim}
\setlog prints a warning message and the same formula:
\begin{verbatim}
***WARNING***: non-finite domain

true
Constraint: Turn is 2*N+1
\end{verbatim}
This means that the formula \emph{might be} satisfiable but  CLP(FD) isn't sure. If we want a more reliable answer we have to bound \Verb+Turn+ or \Verb+N+ to a finite domain:
\begin{verbatim}
N in int(1,5) & Turn is 2*N + 1.
\end{verbatim}
in which case the first solution is:
\begin{verbatim}
N = 1, Turn = 3
\end{verbatim}
and we can get more solutions interactively. On the contrary, if we activate CLP(Q) the finite domain doesn't quite help to get a concrete solution:
\begin{verbatim}
int_solver(clpq).

N in int(1,5) & Turn is 2*N + 1.

true
Constraint: N>=1, N=<5, Turn is 2*N+1
\end{verbatim}

On the other hand, CLP(Q) is complete for linear integer arithmetic while CLP(FD) isn't. This means that if we want to use \setlog to \emph{automatically prove} a property of the program \emph{for all the integer numbers}, we must use CLP(Q)\footnote{As in general non-linear arithmetic is undecidable it's quite difficult to build a tool capable of automatically proving program properties involving non-linear arithmetic.}. Given that simulations don't prove properties it's reasonable to use CLP(FD).

\subsection{Symbolic simulations}
The symbolic execution of a program means to execute it providing to it variables as inputs instead of constants. This means that the executing engine should be able to symbolically operate with variables in order to compute program states as the execution moves forward. As a symbolic execution operates with variables, it can show more general properties of the program than when this is run with constants as input.

\setlog is able to symbolically simulate prototypes, within certain limits. These limits are given by set theory and non-recursive clauses. The following are the conditions under which \setlog can run symbolic simulations\footnote{This is an informal description and not entirely accurate of the conditions for \setlog being able to perform symbolic simulations. These conditions are more or less complex and quite technical.}:
\begin{enumerate}
  \item Recursive clauses are not allowed (for instance \Verb+sum_da+ shown in Section \ref{da}).
  \item Only the operators of Tables \ref{t:setoper} and \ref{t:reloper} are allowed. If the \setlog program uses the cardinality operator (\Verb+size+), the program can't use the operators of Table \ref{t:reloper}. The \Verb+size+ operator is complete only when combined with the operators of Table \ref{t:setoper}.
  \item All the arithmetic formulas are linear\footnote{More precisely, all the integer expressions must be sums or subtractions of terms of the form \Verb+x*y+ with \Verb+x+ or \Verb+y+ constants. All arithmetic relational operators are allowed, even \Verb+neq+.}. 
\end{enumerate}
This means the \setlog code can't use operators of Table \ref{t:listoper} if symbolic simulations are to be done\footnote{The problem with the operators of Table \ref{t:listoper} is that they depend on certain aspects of set theory that aren't implemented in \setlog, yet.}. Actually, many symbolic simulations are still possible even if the above conditions aren't met.

The \setlog prototype of the birthday book falls within the limits of what \setlog can symbolically simulate. For example, starting from the initial state we can call \Verb+addBirthday+ using just variables:
\begin{Verbatim}
birthdayBookInit(K,B) & addBirthday(K,B,N,C,K_,B_).
\end{Verbatim}
in which case \setlog answers:
\begin{Verbatim}
K = {},  
B = {},  
K_ = {N},  
B_ = {[N,C]}
\end{Verbatim}
which is a representation of the expected results. Now we can chain a second invocation to \Verb+addBirthday+ using other input variables:
\begin{Verbatim}
birthdayBookInit(K,B) & addBirthday(K,B,N,C,K1,B1) & addBirthday(K1,B1,M,D,K_,B_).
\end{Verbatim}
in which case the first solution returned by \setlog is:
\begin{Verbatim}
K = {},  
B = {},  
K1 = {N},  
B1 = {[N,C]},  
K_ = {N,M},  
B_ = {[N,C],[M,D]}
Constraint: N neq M
\end{Verbatim}
where we can see a section of the answer which has never appeared before. Indeed, the most general solution that can be returned by \setlog consists of two parts: a (possibly empty) list of equalities between variables and terms (or expressions); and a (possibly empty) list of \emph{constraints}. Each constraint is a \setlog predicate; the returned constraints appear after the word \Verb+Constraint+. The conjunction of all these constraints is always satisfiable (in general the solution is obtained by substituting the variables of type set by the empty set). In this example, clearly, the second invocation to \Verb+addBirthday+ can add the pair \Verb+[M,D]+ to the birthday bool if and only if \Verb+M nin {N}+ (i.e. $M \notin \{N\}$), which holds if and only if \Verb+M+ is different from  \Verb+N+.

\setlog returns a second solution to this symbolic execution:
\begin{Verbatim}
K = {},  
B = {},  
K1 = {N},  
B1 = {[N,C]},  
M = N,  
K_ = {N},  
B_ = {[N,C]}
\end{Verbatim}
produced after considering that \Verb+N+ and \Verb+M+ are equal in which case the second invocation to \Verb+addBirthday+ goes through the branch of \Verb+nameAlreadyExists+ and so \Verb+K_+ and \Verb+B_+ are equal to \Verb+K1+ and \Verb+B1+, which is the expected result as well.

Clearly, symbolic executions allows us to draw more general conclusions about the behavior of the prototype. The next example illustrates what we just said:
\begin{Verbatim}
birthdayBookInit(K,B) & addBirthday(K,B,N,C,K1,B1) &
addBirthday(K1,B1,M,D,K2,B2) & findBirthday(K2,B2,W,X,K2,B2).
\end{Verbatim}
given that \setlog will consider several particular cases regarding whether \Verb+M+, \Verb+N+ and \Verb+W+ are equal or not. For example, the following are the first three solutions returned by \setlog:
\begin{Verbatim}
K = {},  
B = {},  
K1 = {N},  
B1 = {[N,C]},  
K2 = {N,M},  
B2 = {[N,C],[M,D]},  
W = N,  
X = C
Constraint: N neq M

Another solution?  (y/n)
K = {},  
B = {},  
K1 = {N},  
B1 = {[N,C]},  
K2 = {N,M},  
B2 = {[N,C],[M,D]},  
W = M,  
X = D
Constraint: N neq M

Another solution?  (y/n)
K = {},  
B = {},  
K1 = {N},  
B1 = {[N,C]},  
K2 = {N,M},  
B2 = {[N,C],[M,D]}
Constraint: N neq M, N neq W, M neq W
\end{Verbatim}
In the first case \Verb+W = N+ is considered and so \Verb+X+ must be equal to \Verb+C+; the second case is similar to the first one; and in the third \Verb+W+ isn't \Verb+M+ nor \Verb+N+ and so \Verb+X+ can take any value (is this an error?). \setlog returns more solutions some of which are repeated.

Obviously symbolic simulations may combine variables with constants. In general the less the variables we use the less the number of solutions.

\subsection{Inverse simulations}
Normally, in a simulation the user provides inputs and the model returns the outputs. There are situations in which is interesting to get the inputs from the outputs. This means a sort of an inverse simulation.

\setlog is able to perform inverse simulations within the same limits in which is able to perform symbolic simulations. In fact, a careful reading of the previous section reveals that \setlog doesn't really distinguish input from output variables, nor between before and after states. As a consequence, for \setlog is more or less the same to simulate a prototype by providing values for the input variables or for the output variables; in fact, \setlog is able to simulate a prototype just with variables.

Let's see a very simple inverse simulation where we only give the after state:
\begin{Verbatim}
K_ = {maxi,caro,cami,alvaro} &
B_ = {[maxi,160367],[caro,201166],[cami,290697],[alvaro,110400]} &
addBirthday(K,B,N,C,K_,B_).
\end{Verbatim}
The first solution returned by \setlog is the following:
\begin{Verbatim}
K_ = {maxi,caro,cami,alvaro},  
B_ = {[maxi,160367],[caro,201166],[cami,290697],[alvaro,110400]},  
K = {maxi,caro,cami},  
B = {[maxi,160367],[caro,201166],[cami,290697]},  
N = alvaro,  
C = 110400
\end{Verbatim}

When the Z specification is deterministic, the corresponding \setlog prototype will be deterministic as well. Therefore, for any given input there will be only one solution. However, the inverse simulation of a deterministic prototype may generate a number of solutions. This is the case with the above simulation. The first solution computed by \setlog considered the case where \Verb+N = alvaro+ and \Verb+C = 110400+, but this isn't the only possibility. Then, when we ask \setlog for another solution we get, for instance, the following:
\begin{Verbatim}
K_ = {maxi,caro,cami,alvaro},  
B_ = {[maxi,160367],[caro,201166],[cami,290697],[alvaro,110400]},  
K = {maxi,caro,alvaro},  
B = {[maxi,160367],[caro,201166],[alvaro,110400]},  
N = cami,  
C = 290697
\end{Verbatim}
which means that \Verb+K_+ and \Verb+B_+ may have been generated by starting from some \Verb+K+ and \Verb+B+ where \Verb+cami+'s birthday isn't in the book and so we can add it.

\subsection{\label{evaluacion}Evaluation of predicates}
At the end of Section \ref{simples} we showed how to start a simulation from a state different from the initial state. We also said that this entails some risks as manually writing the start state is error prone which may lead to an unsound state. In this section we will see how to avoid this problem by using a feature of \setlog that is useful for other verification activities, too.

Let's consider the following state of the birthday book:
\begin{Verbatim}
Known = {maxi,caro,cami,alvaro}
Birthday = {[maxi,160367],[caro,201166],[cami,290697],[alvaro,110400]}
\end{Verbatim}
Starting a simulation from this state may give incorrect results if it doesn't verify the state invariant defined for the specification. Recall that the state invariant for the birthday book is:
\begin{schema}{BirthdayBookInv}
BirthdayBook
\where
known = \dom birthday
\end{schema}

With \setlog is very simple to check that \Verb+S0+ verifies the state invariant. We first translate $BirthdayBookInv$ into \setlog:
\begin{Verbatim}
birthdayBookInv(Known,Birthday) :-
  dom(Birthday,Known).
\end{Verbatim}
Afterwards, we invoke the new clause passing in to it the state defined above:
\begin{Verbatim}
Known = {maxi,caro,cami,alvaro} &
Birthday = {[maxi,160367],[caro,201166],[cami,290697],[alvaro,110400]} &
birthdayBookInv(Known,Birthday).
\end{Verbatim}
in which case \setlog returns the values of \Verb+Known+ and \Verb+Birthday+, meaning that \Verb+birthdayBookInv+ is satisfied. If this weren't the case the answer would have been \Verb+no+. If we call \Verb+birthdayBookInv+ passing in to it the values of the variables:
\begin{Verbatim}
birthdayBookInv({maxi,caro,cami,alvaro},
                {[maxi,160367],[caro,201166],[cami,290697],[alvaro,110400]}).
\end{Verbatim}
the answer is \Verb+yes+. 

Now, let's say we have written a state not satisfying the state invariant, for instance (note that \Verb+maxi+ is missing from \Verb+known+):
\begin{Verbatim}
Known = {caro,cami,alvaro} &
Birthday = {[maxi,160367],[caro,201166],[cami,290697],[alvaro,110400]} &
birthdayBookInv(Known,Birthday).
\end{Verbatim}
Then, the answer of \setlog is \Verb+no+. 

In general we can take any property and evaluate it on some values. For example, \Verb+Birthday+ must be a partial function in any given state. Hence, for instance, if we execute:
\begin{Verbatim}
pfun({[maxi,160367],[caro,201166],[cami,290697],[alvaro,110400]}).
\end{Verbatim}
\setlog answers \Verb+yes+ and if we execute:
\begin{Verbatim}
pfun({[maxi,160367],[maxi,201166],[cami,290697],[alvaro,110400]}).
\end{Verbatim}
it answers \Verb+no+.

\section{\label{pruebas}Automated proofs with \setlog}
Evaluating properties with \setlog helps to run correct simulations by checking that the starting state is correctly defined. It also helps to \emph{test} whether or not certain properties are true of the specification or not. However, it would be better if we could  \emph{prove} that these properties are true of the specification. In this section we will see how \setlog allows us to prove that the operations of a specification preserve the state invariant.

So far we have used \setlog as a (prototype) programming language. However, \setlog is also a \emph{satisfiability solver}. This means that \setlog is a program that can decide if formulas of some theory are \emph{satisfiable} or not. In this case the theory is the theory of finite sets and binary relations combined with the operators listed in Tables \ref{t:setoper} and \ref{t:reloper} combined with the linear integer arithmetic\footnote{In what follows we will only mention the theory of finite sets but the same is valid for this theory combined with linear integer algebra.}. 

If $F$ is a formula depending on a variable, we say that $F$ is \emph{satisfiable} if and only if:
\[
\exists y: F(y)
\]

In the case of \setlog, $y$ is quantified over \emph{all} finite sets. Therefore, if \setlog answers that $F$ is satisfiable it means that there exists a finite set satisfying it. Symmetrically, if \setlog says that $F$ isn't satisfiable it means that there is no finite set satisfying it. Formally, $F$ is an unsatisfiable formula if:
\begin{equation}\label{eq:unsat}
\forall y: \lnot F(y)
\end{equation}
If we call $G(x) \defs \lnot F(x)$ then \eqref{eq:unsat} becomes:
\begin{equation}\label{eq:sat}
\forall y: G(y)  
\end{equation}
which means that $G$ is true of every finite set. Putting it in another way, $G$ is \emph{valid} with respect to the theory of finite sets; or, equivalently, $G$ is a \emph{theorem} of the theory of finite sets.

\vspace{3mm}
\colorbox{gray!20}{%
\parbox{.92\textwidth}{%
In summary, if \setlog decides that \emph{$F$ is unsatisfiable}, then we know that \emph{$\lnot F$ is a theorem}.
}}
\vspace{3mm}

In other words, \eqref{eq:unsat} and \eqref{eq:sat} are two sides of the same coin: \eqref{eq:unsat} says that $F$ is unsatisfiable and \eqref{eq:sat} says that $G$ (i.e. $\lnot F$) is a theorem.

\vspace{3mm}
\colorbox{gray!20}{%
\parbox{.92\textwidth}{%
Recall to activate CLP(Q) when performing proofs.
}}
\vspace{3mm}

Recall that in order to prove that an operation, $T$, of a Z specification preserves the state invariant $I$ we have to discharge the following proof obligation:
\begin{equation}\label{e:inv}
I \land T \implies I'
\end{equation}
If we want to use \setlog to discharge \eqref{e:inv} we have to ask \setlog to check if the negation of \eqref{e:inv} is \emph{unsatisfiable}. In fact, we need to execute the following \setlog \emph{program}:
\begin{equation}\label{e:neginv}
I \land T \land \lnot I'
\end{equation}
because $\lnot(I \land T \implies I') \equiv \lnot(\lnot(I \land T) \lor I') \equiv I \land T \land \lnot I'$.

The problem in using \setlog to decide the satisfiability of \eqref{e:neginv}, is the predicate $\lnot I'$. More precisely, we shouldn't use logical negation because if we do, in general, \setlog won't be able to prove that the formula is unsatisfiable. In any case, logical negation can be avoided in \setlog \emph{as long as we work with the operators of Tables \ref{t:setoper} and \ref{t:reloper} and linear integer algebra}. Recall that for each operator of Tables \ref{t:setoper} and \ref{t:reloper} \setlog implements its negation. For example, instead of $\lnot x \in A$ we should write $x \notin A$; instead of $\lnot A = B$, we should write $A \neq B$; instead of $\lnot pfun(f)$ we should write $npfun(f)$; instead of $\lnot X > Y$ we should write $X \leq Y$; and so forth and so on. Then, coming back to formula \eqref{e:neginv}, if $I'$ is a compound predicate (i.e. a predicate formed by conjunction, disjunction, implication, etc.), then we must distribute the negation all the way down to the atoms at which point we use the negations of the operators of Tables \ref{t:setoper} and \ref{t:reloper}.

\bigskip

In the Z specification of the birthday book we proposed as invariant the following formula:
\[
\dom birthday = known
\]
Hence, if we want to prove that $AddBirthday$ preserves this invariant, we have to prove the following:
\begin{equation}\label{eq:AddBirthdayk=b}
\dom birthday = known \land AddBirthday \implies \dom birthday' = known'  
\end{equation}
However, it we're going to use \setlog to prove \eqref{eq:AddBirthdayk=b} we need to see if  the translation into \setlog of the following formula is unsatisfiable (due to \eqref{e:neginv}):
\[
\dom birthday = known \land AddBirthday \land \dom birthday' \neq known'
\]
The translation into \setlog of that formula is:
\begin{Verbatim}
dom(B,K) &
addBirthday(K,B,N,C,K_,B_) &
ndom(B_,K_).
\end{Verbatim}
When \setlog is called to solve this formula (or to run this program) the answer is \Verb+no+ because \setlog finds it unsatisfiable. If this formula is unsatisfiable, then its negation is a theorem. And the negation of this \setlog formula is \eqref{eq:AddBirthdayk=b}. Hence, if \setlog answers \Verb+no+ we know that $AddBirthday$ preserves the state invariant.

\bigskip

So we have automatically proved that $AddBirthday$ preserves this invariant for every \emph{finite} set. From this, we can't conclude that $AddBirthday$ preserves this invariant for every set but, will the implementation of the birthday book ever process an \emph{infinite} set of persons? The reader can try to prove that the other operations of the birthday book preserve this invariant.

\bigskip

When the Z specification of the birthday book was translated into \setlog, we mentioned that when translating $birthday$ its type at the \setlog level is \Verb+stype([name,date])+---which is interpreted as the Z type $NAME \rel DATE$. Besides we said we didn't encode $birthday$ as a function because in \setlog functions aren't a type (as they are in, for instance, functional programming languages). Finally, we said that we will use \setlog to prove that  $birthday$ is a function in every state, i.e. it's a state invariant. In other words, with \setlog's type system we can express that $birthday$ is a binary relation between $NAME$ and $DATE$; and then we can use \setlog to prove that actually $birthday$ is a function. For more details reread Section \ref{codigosetlog}; in particular we suggest to take a look at shcema $BirthdayBookInv$.

Then, we will prove that:
\begin{equation}\label{eq:AddBirthdaypfun}
birthday \in \_\pfun\_ \land AddBirthday \implies birthday' \in \_\pfun\_  
\end{equation}
where $\_\pfun\_$ denotes the set of all partial functions (with any domain and range). Again, as we will use \setlog to prove that, we need to negate the formula:
\[
birthday \in \_\pfun\_ \land AddBirthday \land birthday' \notin \_\pfun\_
\]
where we interpret $birthday' \notin \_\pfun\_$ as $birthday'$ isn't a partial function. Now,  the corresponding \setlog formula is:
\begin{Verbatim}
pfun(B) &
addBirthday(K,B,N,C,K_,B_) &
npfun(B_).
\end{Verbatim}
where we translate $birthday' \notin \_\pfun\_$ as \Verb+npfun(B_)+ because this predicate states that its argument isn't a partial function (see Table \ref{t:reloper}). When \setlog is asked to solve the above formula (or to run the above program) the answer is the following:
\begin{Verbatim}
B = {[N,_N2]/_N1},  
K_ = {N/K},  
B_ = {[N,_N2],[N,C]/_N1}
Constraint: pfun(_N1), comppf({[N,N]},_N1,{}), N nin K, set(_N1), [N,C] nin _N1, 
            C neq _N2, [N,_N2]nin _N1, set(K)
\end{Verbatim}

\emph{What? Weren't we expecting \Verb+no+ for an answer? The whole idea is that $AddBirthday$ does preserve the fact that $birthday$ is a function, isn't it? Then, what does it mean that \setlog returned a solution?} Clearly, it means that \eqref{eq:AddBirthdaypfun} isn't a theorem. Why \eqref{eq:AddBirthdaypfun} isn't a theorem (we were pretty sure of that)? The answer returned by \setlog helps us to understand the problem. Take a look at it and observe that \Verb+B = {[N,_N2]/_N1}+ but it doesn't say that \Verb+N+ belongs to \Verb+K+ when, as we proved in \eqref{eq:AddBirthdayk=b}, \Verb+K+ is equal to the domain of \Verb+B+ (and \Verb+N+ clearly belongs to the domain of \Verb+B+). The problem is that $AddBirthday$ states that $n? \notin known$ instead of $n? \notin \dom birthday$. As we know that $known = \dom birthday$  then $n? \notin known$ and $n? \notin \dom birthday$ are equivalent. But $known = \dom birthday$ is not part of \eqref{eq:AddBirthdaypfun}, and so \setlog can't use this equality in the proof. Given that we have \emph{already} proved that $known = \dom birthday$ is a state invariant we can \emph{assume} it holds in \eqref{eq:AddBirthdaypfun}. Then, we now call \setlog to solve the following formula:
\begin{Verbatim}
dom(B,K) &                           %%% hypothesis
pfun(B) &
addBirthday(K,B,N,C,K_,B_) &
npfun(B_).
\end{Verbatim}
and now \setlog answers \Verb+no+.

Recall that we don't need to prove that the domain of $birthday$ is a subset of $NAME$ and that its range is a subset of $DATE$ because this is guaranteed by the typechecker.

In addition of proving that each operation of the specification \emph{preserves} the state invariant we must prove that the initial state \emph{satisfies} it (because otherwise the system would start from an unsound state and thus the whole system would be at danger). In general, this proof is simpler. We start by proving that in the initial state $\dom birthday = known$ is satisfied:
\begin{Verbatim}
birthdayBookInit(K,B) &
ndom(B,K).
\end{Verbatim}
We follow by proving that in the initial state $birthday$ is a partial function:
\begin{Verbatim}
birthdayBookInit(K,B) &
npfun(B).
\end{Verbatim}
And we finish by proving that the initial domain of $birthday$ is a subset of $NAME$:
\begin{Verbatim}
birthdayBookInit(K,B) &
dom(B,DB) & ran(B,RB) &
(nsubset(DB,NAME) or nsubset(RB,DATE)).
\end{Verbatim}

\subsection{Typechecking and proofs}
As with simulations, proofs can be done with or without the typechecker activated. All we did above assumes the typechecker isn't active (command \Verb+notype_check+). Activating the typechecker makes proofs somewhat more cumbersome (as users need to give the type of all variables) but they are safer (because the typechecker can catch errors that \setlog's deductive system can't). 

Given that from a verification perspective proofs are more critical than simulations, it is recommended to activate the typechecker before discharging the invariance lemmas.

We will show, then, how to run proofs when the typechecker is active (command \Verb+type_check+). If we run the proof without declaring the types of variables:
\begin{Verbatim}
dom(B,K) &
addBirthday(K,B,N,C,K_,B_) &
ndom(B_,K_).
\end{Verbatim}
we get a type error:
\begin{Verbatim}
***ERROR***: type error: variable B has no type declaration
\end{Verbatim}
Hence, we have to give the type for each variable by means of predicate \Verb+dec+:
\begin{Verbatim}
dec([B,B_],bb) & dec([K,K_],kn) & dec(N,name) & dec(C,date) &
dom(B,K) &
addBirthday(K,B,N,C,K_,B_) &
ndom(B_,K_).
\end{Verbatim}
in which case we get a \Verb+no+ answer. Clearly, if types are wrong:
\begin{Verbatim}
dec([B,B_],bb) & dec([K,K_],kn) & dec(N,date) & dec(C,name) &
dom(B,K) &
addBirthday(K,B,N,C,K_,B_) &
ndom(B_,K_).
\end{Verbatim}
we get a type error:
\begin{Verbatim}
***ERROR***: type error: 
  in addBirthday(K,B,N,C,K_,B_) arguments have the wrong type:
    N is date but should be name
    C is name but should be date
\end{Verbatim}

\bibliographystyle{eptcs}
\bibliography{/home/mcristia/escritos/biblio}

\appendix

\section{\label{ap:bb}The \setlog implementation of the birthday book}

\begin{Verbatim}
:- dec_type(bb,stype([name,date])).
:- dec_type(kn,stype(name)).

:- dec_p_type(birthdayBookInit(kn,bb)).
birthdayBookInit(Known,Birthday) :-
  Known = {} &
  Birthday = {}.

:- dec_p_type(birthdayBookInv(kn,bb)).
birthdayBookInv(Known,Birthday) :-
  dom(Birthday,Known).

:- dec_p_type(addBirthdayOk(kn,bb,name,date,kn,bb)).
addBirthdayOk(Known,Birthday,Name_i,Date_i,Known_,Birthday_) :-
  Name_i nin Known &
  un(Known,{Name_i},Known_) &
  un(Birthday,{[Name_i,Date_i]},Birthday_).

:- dec_p_type(nameAlreadyExists(kn,bb,name,kn,bb)).
nameAlreadyExists(Known,Birthday,Name_i,Known_,Birthday_) :-
  Name_i in Known &
  Known_ = Known &
  Birthday_ = Birthday.

:- dec_p_type(addBirthday(kn,bb,name,date,kn,bb)).
addBirthday(Known,Birthday,Name_i,Date_i,Known_,Birthday_) :-
  addBirthdayOk(Known,Birthday,Name_i,Date_i,Known_,Birthday_)
  or
  nameAlreadyExists(Known,Birthday,Name_i,Known_,Birthday_).

:- dec_p_type(findBirthdayOk(kn,bb,name,date,kn,bb)).
findBirthdayOk(Known,Birthday,Name_i,Date_o,Known_,Birthday_) :-
  Name_i in Known & 
  apply(Birthday,Name_i,Date_o) &
  Known_ = Known &
  Birthday_ = Birthday.

:- dec_p_type(notAFriend(kn,bb,name,kn,bb)).
notAFriend(Known,Birthday,Name_i,Known_,Birthday_) :-
  Name_i nin Known & 
  Known_ = Known &
  Birthday_ = Birthday.

:- dec_p_type(findBirthday(kn,bb,name,date,kn,bb)).
findBirthday(Known,Birthday,Name_i,Date_o,Known_,Birthday_) :-
  findBirthdayOk(Known,Birthday,Name_i,Date_o,Known_,Birthday_)
  or
  notAFriend(Known,Birthday,Name_i,Known_,Birthday_).

:- dec_p_type(remind(kn,bb,date,kn,kn,bb)).
remind(Known,Birthday,Today_i,Cards_o,Known_,Birthday_) :-
  rres(Birthday,{Today_i},M) & dec(M,bb) &
  dom(M,Cards_o) &
  Known_ = Known &
  Birthday_ = Birthday.
\end{Verbatim}
\end{document}